\documentclass[acmlarge,screen,authorversion,nonacm]{acmart}

\copyrightyear{2021}

\usepackage{tabularx}
\usepackage{amsmath}
\usepackage{balance}
\usepackage{multirow}
\usepackage{threeparttable}
\usepackage{fancyhdr}

\makeatletter
\let\@authorsaddresses\@empty
\makeatother

\begin{document}

\title{A Need-finding Study for Understanding Text Entry in Smartphone App Usage}

\author{Toby Jia-Jun Li}
\email{toby.j.li@nd.edu}
\authornote{Work done at Carnegie Mellon University}
\affiliation{%
  \institution{University of Notre Dame \& Carnegie Mellon University}
  \city{Notre Dame}
  \state{IN}
}

\author{Brad A. Myers}
\email{bam@cs.cmu.edu}
\affiliation{%
  \institution{Carnegie Mellon University}
  \city{Pittsburgh}
  \state{PA}
}

\begin{abstract}
 Text entry makes up about one-fourth of the smartphone interaction events, and is known to be challenging and difficult. However, there has been little study about the characteristics of text entry in the context of smartphone app usage. In this paper, we present a mixed-method in-situ study conducted in 2016 with 17 active smartphone users to better understand text entry in smartphone app usage. Our results show 80\% of text was entered into communication apps, with different apps exhibiting distinct usage patterns. We found that structured data such as URLs and email addresses are rarely typed but instead are auto-completed or replaced with search, copy-and-paste is rarely used, and sessions of smartphone usage with text entry involve more apps and last longer. We conclude with a discussion about the implications on the development of systems to better support mobile interaction.    
\end{abstract}

\maketitle
\thispagestyle{fancy}
 
\addtolength{\footskip}{2.0\baselineskip}%
\fancyfoot[L]{\textit{\textbf{Unpublished Preprint --- Copyright held by the owner/author(s)}}}%
\fancyhead[OR]{\thepage}%
\fancyhead[EL]{\thepage}%

\section{Introduction}
Smartphones are the primary devices on which users send and receive messages, take notes, and search for information on the go. Text entry is an important aspect of users' mobile interactions and makes up about one-fourth of the total interaction events according to our data (with the other three-quarters being taps, swipes, focuses and other interactions). It is also long been known that text entry is a particularly challenging and difficult activity in using mobile phones~\cite{kristensson_grand_2013, mackenzie_text_2002, vertanen_invisid_2016}. 

Many solutions have been proposed to improve the performance and user experience of mobile text entry, including novel (virtual) keyboards (e.g.,~\cite{zhai_performance_2002}), alternative non-keyboard text entry techniques (e.g.,~\cite{wobbrock_edgewrite:_2003}), text prediction (e.g.,~\cite{vertanen_velocitap:_2015}), and apps to support a specific domain of text entry (e.g.,~\cite{chang_improving_2013}). There also have been text corpora which have been collected from a single domain of smartphone usage, such as for email~\cite{vertanen_versatile_2011}  or chat rooms~\cite{zhai_performance_2002}, which have been used to help evaluate text entry methods. However, surprisingly little study has focused on understanding users' text entry in the context of smartphone app usage. This lack of knowledge makes it hard to properly evaluate if a proposed solution to text entry problems is relevant. It also hinders the development of new text entry mechanisms, as researchers often need to make guesses or assumptions about users' text entry behaviors.

To help fill in this gap, we conducted a mixed-method study including surveys, interviews, diaries and in-situ logging to characterize the users' mobile text entry behaviors. In our study, we recorded 1,607,199 total events, of which 401,875 were text-editing events, representing 476,322 characters entered into 155,300 app usages of 364 unique apps from 17 active smartphone users (mostly university students) over 2 weeks. We also gathered the participants' reflections on their usages through interviews and diaries.

Using the collected data, in this paper, we mainly focus on answering the following research questions:
\begin{itemize}
\item[\textbf{RQ1:}] To which applications are data entered into smartphones?

\item[\textbf{RQ2:}] What kinds of textual data are entered into smartphones?

\item[\textbf{RQ3:}] Through what means are data entered (typing vs. copy-and-paste vs. auto-completion)?

\item[\textbf{RQ4:}] How are text-entering sessions different from the non-text entering sessions?

\item[\textbf{RQ5:}] How is text transferred among apps within sessions?
\end{itemize}

In summary for the above questions, we found that for our users, over 80\% of text was entered into communication apps like instant messaging (IM), SMS and email. Among those apps, the participants exhibit different text entry patterns: only a small portion of email app usages involve text entry compared to IM and SMS, but more email app usage involves heavy text entry (more than 100 characters) than the other two communication apps. 

Only a small portion (2.5\%) of the strings entered into the smartphones by our participants includes structured data in the form of phone numbers, email addresses, URLs or street addresses, and most of them go to where one might expect (e.g., URLs are mostly entered in a browser). In addition, for about half of the times that the clipboard was used, the clipboard contents contained structured data, mostly URLs. 

We found that in our data, 60\% of the text was typed character by character into phones. Most of the rest were auto-completed, with less than 1\% being entered through copy-and-paste. Looking at the various sessions, we found that sessions with text entry on average last longer and have more apps used than those with no text entry. Our interview results showed that most of our participants found copy-and-paste difficult to use and preferred alternative ways of sharing data, such as using the Android ``share to'' API, or taking a screenshot of an app and sharing it with other people. 

While our study of smartphone usage behavior was performed on a sub-population of users for a limited period of time, and the results will not necessarily generalize across all different populations, our findings provide value, as discussed in~\cite{church_understanding_2015}, in uncovering interesting user practices and providing insights into how future smartphone interactions and apps can better support text entry and thus improve the overall user experience.

\section{Related Work}
There have been prior studies that tried to understand various aspects of smartphone usage. Böhmer et al.~\cite{bohmer_falling_2011} ran a large-scale logging study of app launches and found that the use of an app on average lasts less than a minute. They also discovered that communication is the most often used app category, with some app categories being used more often at certain times. The LiveLab project~\cite{shepard_livelab:_2011} also collected a large dataset of smartphone app usage records. Many prior studies (e.g.,~\cite{barkhuus_empowerment_2011, brown_iphone_2013, church_understanding_2011, karlson_mobile_2010, teevan_understanding_2011}) investigated how smartphone usage is triggered by the context and how smartphone interactions can be integrated with the user's activity. Karlson et al.~\cite{karlson_working_2009} looked at smartphone use in the context of task flow, suggesting that interruptions affect mobile productivity and found it common for users to move to a PC to complete a task started on the phone. Some studies also focused on characterizing users' information needs: Sohn et al.~\cite{sohn_diary_2008} conducted a diary study and discovered that people use diverse and sometimes ingenious ways to obtain information on smartphones. 

To fully understand smartphone usage, it is not sufficient to only analyze the use of individual apps, as each smartphone app is designed to handle a limited number of domains (usually one), which is not sufficient to support many cross-domain tasks~\cite{sun_understanding_2015, sun_intelligent_2016}. Thus, many researchers have looked at multi-app usage sessions.  For instance, for mobile search tasks, Carrascal and Church~\cite{carrascal_-situ_2015} reported that users frequently switch among mobile search apps and other apps, and that sessions with mobile searches also tend to have more apps used, have a longer duration; and use some categories of apps more intensively. On the other hand, many smartphone usage sessions can be very short with studies reporting that 40\% of sessions are shorter than 15 seconds~\cite{ferreira_contextual_2014} and 38\% contain no app usage but only interactions with the lock screen or launcher~\cite{banovic_proactivetasks:_2014} (for example, to look at notifications or the time). Jesdabodi and Maalej~\cite{jesdabodi_understanding_2015} proposed the concept of ``usage states'' to categorize usage sessions based on the topic of user's task. Many projects (e.g.,~\cite{jones_revisitation_2015, liao_feature_2013, parate_practical_2013, shin_visual_2012, srinivasan_mobileminer:_2014, xu_preference_2013, zou_prophet:_2013}) investigated the ``app chain''---the temporal patterns in app launches or the contextual information of the smartphone usage to predict the next app to be launched. Sequences of app screens found in user interaction traces may also be used to represent the user intents of mobile tasks~\cite{li_screen2vec_2021}. Systems like \textsc{Helper}~\cite{sun_helpr:_2016} and \textsc{Sugilite}~\cite{li_sugilite:_2017, li_pumice:_2019, li_interactive:_2020} proposed using intelligent agents to automate tasks across multiple apps and domains.  

However, none of the above studies looked at the text entry aspect of smartphone usage, nor is there any dataset for text entry behaviors in app usage available to the community. There were text corpora in some prior projects~\cite{stocky_commonsense_2004, vertanen_versatile_2011, zhai_performance_2002} collected from users' mobile text entry in a single domain (like emails or chat rooms) for the purpose of training and evaluating the performance of text entry methods. But they did not consider text entry behaviors in the context of multiple mobile app interactions, nor did they characterize the text entry behaviors so we can learn what data do users enter or what apps are these data entered into. There are also studies on the usage of a single domain of app with heavy text entry like instant messaging~\cite{church_whats_2013, dingler_ill_2015} and SMS~\cite{battestini_large_2010}.

Various systems have been developed to better support mobile text entry. Some~\cite{chang_plug-architecture_2013, chang_improving_2013, nardi_collaborative_1998} support easier and more efficient entry and processing of structured data like dates, addresses and URLs. Fuccella et al.'s work~\cite{fuccella_gestures_2013} supports mobile text editing with gestures. Although there are many solutions proposed, there are few studies about how often those solutions will be relevant, or how prominent the reported problems are, as we do not yet know in which apps do users enter and process structured data, nor do we know how those data are entered and what types of structured data are common. We hope our results reported here can help the community to have a deeper understanding of users' text entry behaviors with respect to app usage and usage sessions, and provide resources for the development of future systems to better support smartphone text entry.

\section{Pre-Survey}
To help us better design our studies, we conducted an online survey using SurveyMonkey about the users' text entry and cross-app usage on their smartphones. By posting the link on Facebook groups and to mailing lists at Carnegie Mellon University, we collected 65 responses between December 2015 and January 2016. 70\% of the respondents were between 18--24 years old, 87\% were students, 57\% used iPhone while the rest (43\%) used Android phones. In the survey, we asked the respondents to estimate their frequency of entering text for different purposes on their smartphones, and to describe the tasks, the goals, the procedures and the apps used for each of their most recent text entry situations. We also asked about their copy-and-paste usage, retyping, cross-application interactions and repetitive sequences of actions. 

When asked about \textit{which} applications they most enter data into on their smartphones, 76\% of the respondents self-reported that they often entered data into map/navigation apps, 70\% said they often entered data into instant messaging apps, and 66\% said search engines. On the other end of the spectrum, most people reported that they never or seldom entered data into spreadsheets (58\% of the respondents said never; 30\% said seldom), text editing apps (30\% never; 50\% seldom) or note-taking apps (10\% never; 53\% seldom) on their smartphones.

For copy-and-paste, 3\% respondents reported that they never used copy and paste, 12\% seldom used it, 52\% some-times used it and 33\% often do so. Among the 53 responses who reported about their most recent copy-and-paste usage, the contents reported by 25 people (47\%) consisted of ``structured text,'' which we defined in the survey as URLs, phone numbers, email addresses and street addresses. The most popular type of structured text was URL, which was reported by 15 people (28\%). Respondents also shared their most recent case of retyping contents instead of using copy-and-paste, mostly because they said that the text field did not allow copy-and-paste, or that copy-and-paste is too hard to use. For example, as a rationale for not using copy-and-paste, some of our respondents wrote:

\begin{quote}
\textit{``I saw it [the text] on a link and didn't want to accidentally click on the link.'' }

\textit{``… It was that either it was ineffective to try to copy paste due to resolution, clutter of other text in the browser or the address as typed didn't work on Google Maps.'' }

\textit{``The input box doesn't support paste''.}
\end{quote}

The respondents also shared their most recent experiences of using multiple apps for a single task. Sample usage scenarios they described included:

\begin{quote}
\textit{``Plan the event using email or messenger and check whether I'm free using Calendar.'' ``[I used] Airbnb for the address, looked up the route through Google Maps, checked the bus schedule with [the] browser.''}

\textit{``I opened Gmail to locate the receipt, I noted the total amount I had spent and divided it fairly between my friends (evenly, in this case). I then opened Venmo and charged each friend who owed me separately.''}

\textit{``Find the name in Yelp, check rating and reviews, and then find the estimated time it takes from my place to the bar by using Google Maps.''}
\end{quote}

31 respondents were able to describe their most recent experiences of performing a repetitive task. A respondent talked about how she kept all her expenses in memos and she ``saw the money I spent each day on notes and then added each entry on calculator.'' One other talked about how he needed to switch back-and-forth between the Cisco tool, system settings and the Browser to connect to the office network every morning.

The results of the pre-survey further suggest that users exhibit different text entry behaviors in different domains, and the amount of text, the frequency of text entry and the type of content entered vary for different categories of apps. It also suggests that in many situations, users need to use multiple apps, sometimes with data being transferred among the apps, to perform a single task. The outcome of the pre-survey shows the need for a more detailed empirical study to better understand users' text entry and cross-application behaviors, which motivates the main study we will talk about in the following sections. 

\section{Longitudinal Study}
We conducted a two-week longitudinal study with 17 active smartphone users in February and March of 2016 to understand the users' data entry and cross-application usage in depth. The study consisted of two in-lab user interviews and a 2-week-long in-situ logging and diary study. We collected detailed logs of the users' smartphone interactions (clicking, tapping etc.), data entry behaviors (keystroke, copy-and paste), app usage and the corresponding usage context using a tracking tool that we created and installed on their phones. Through surveys and interviews, we also gathered the users' reflections on their smartphone interactions and descriptions of their specific usage sessions.

\subsection{Participants}
We recruited 17 participants (7 males and 10 females, ages 18--45, mean age 25.2, SD=5.48) from Carnegie Mellon University. All participants were proficient in English and used English as the main language in their interactions with smartphones. We recruited participants who use Android phones running Android 4.4 or above as their main mobile phones and self-identified as active smartphone users. During the course of study, the participants on average spent 4.2 hours on their smartphones per day, with the least active participant spending an average of 1.1 hours per day and the most active one spending on average 7.7 hours per day, so it seems reasonable to consider all of them to be active users. Each participant signed the informed consent form, and received a compensation of \$50.

\subsection{Tracking Tool}
We implemented an Android app named \textsc{CrossAppTracker} to track the participants' smartphone interactions. \textsc{CrossAppTracker} runs as an Android accessibility service and records all of the participants' interactions with the phone UI across all smartphone apps. Each interaction (tapping, clicking, keystroke, focusing, clipboard event and screen state change) is recorded as a tracked event. Tracked events include when the user clicks or focuses on a UI element, enters or edits some text, when the window state changes, or when the contents of the clipboard changes. For each tracked event, the tracker also logs the name of the current app, the name of the current Android activity\footnote{An activity is a single, focused thing that the user can do~\cite{google_android_2016} within an Android application. Each activity often represents a single screen with a user interface.}, the text or content description\footnote{Content description is an accessible text label for a UI element that can be read out by a screen reader.} of the UI element, a unique identifier of the UI element, a timestamp and the identifier (SSID) of the connected Wi-Fi network. In addition to these events, \textsc{CrossAppTracker} also records an event whenever the screen was turned on or off. \textsc{CrossAppTracker} automatically uploads the logs every 8 hours to our server, so the experimenter could look for interesting interactions. If the upload failed, \textsc{CrossAppTracker} would retry later to ensure the completeness of the log.

App GUI screens and the user interaction traces on these app GUI screens often contain sensitive personal information~\cite{li_privacy:_2020}. To protect privacy, \textsc{CrossAppTracker} allows the participants to specify a list of apps to exclude from keystroke logging, for example those that contain sensitive data, like banking apps and medical apps. It also avoids logging any data marked as encrypted or entered into password fields. Still, the logs do contain sensitive data, so they are kept in a protected place and processed in accordance to our IRB-approved study protocol. 

A limitation of \textsc{CrossAppTracker} is that it can only listen to interaction events in native Android apps, but not when the user uses a web application. However, we do not expect this limitation to have a major impact on our results, as all but one of our 17 participants reported that they would almost always prefer using a native Android app to a web app when both are available. A study from Yahoo also reported that 90\% of users' time on smartphones is spent in apps, while only 10\% is spent in the browser~\cite{khalaf_seven_nodate}.

\subsection{Study Procedure}
The study consisted of four main components:

\subsubsection{Pre-interview}
In 30-minute semi-structured interview sessions, the participants were asked to reflect on their smartphone usage in a one-on-one lab interview setting. They described the motivations, goals and procedures of concrete examples of their smartphone tasks that are cross-app, text entry heavy or repetitive. They also discussed the apps they used the most, how they split and coordinate tasks between their smartphone and other computing devices (computers, tablets, wearables), how they interact with the smartphone's virtual assistants (e.g., Google Now, S Voice), how they use copy-and-paste on their phones, and any smartphone tasks they find frustrating to perform, or would like to automate.

\subsubsection{In-situ Logging}
During the pre-interview, we installed \textsc{CrossAppTracker} onto the participants' smartphones. \textsc{CrossAppTracker} recorded all of the participants' smartphone interactions as discussed above, and uploaded the logs to our server, so the experimenter could look for interesting interactions and ask the participants to reflect on those in their daily diaries.

Each participant was asked to keep \textsc{CrossAppTracker} running in the background at all times for at least 14 days while using their smartphones as they normally would. Some participants recorded more than 14 days of data due to the scheduling of interview sessions, since the \textsc{CrossAppTracker} software was removed at the exit interview.  

\subsubsection{Diary Study}
At the end of each day during the 2-week in-situ study, each participant filled out an online questionnaire about their smartphone usage on that day. Each questionnaire consisted of 4 questions. The first 3 questions were constant. The last one was a personalized question made by the experimenter about a specific usage session from that day. The 3 constant questions asked the respondents to describe the goals, procedures and contexts of their text entry heavy, cross-app or repetitive tasks on that day, if any. The personalized question was often in the form ``we saw you used [list of apps] at [time] today, can you report what was the task and how did you use the above apps to perform the task?'' 

\subsubsection{Exit Interview}
Each of the participants came into our lab for a 30-minute semi-structured interview session after completing the in-situ logging and diary study. During the session, the participants had their \textsc{CrossAppTracker} removed, and they were asked to clarify any confusing entries in their diaries and reflect on their smartphone usage of the past two weeks. 

\subsection{Dataset}
The raw dataset we collected from the 17 participants through the in-situ logging contained 1,607,199 tracked events, representing 155,300 application usages in 37,496 sessions. An application usage contains all the tracked events starting from when an app became active in the foreground until either another app became active instead, or the phone was locked or turned off. Similarly, a session includes all the application usages starting from when the phone was unlocked or turned on until when the phone was locked or turned off. In the dataset, the 17 participants used 364 unique apps over the course of our study.

\section{Results}
\subsection{Characterizing Text Entry Behaviors in App Usage}
In this section, we mainly try to understand the characteristics of the users' text entry behaviors. More specifically, we attempt to answer three research questions: 
\begin{itemize}

\item[\textbf{RQ1:}] To which applications are data entered into smartphones?

\item[\textbf{RQ2:}] What kinds of data are entered into smartphones?

\item[\textbf{RQ3:}] Through what means are data entered (typing vs. copy-and-paste vs. auto-completion)?

\end{itemize}

Our dataset contained 401,875 text-editing events (addition, deletion or replacement of characters) made by the 17 participants in 348 different applications over roughly 2 weeks, representing a total of 476,322 characters entered, which works out to an average of 28,018 characters per person (SD=20802.5). Excluding symbols and spaces, 357,306 characters make up 71,974 words, with an average word length 4.96. All the characters constitute 12,504 separate strings entered, or 52.5 strings per participant per day. 

We group all the text-editing events into the corresponding app that they belong to, and use the app usage as our unit of analysis. Among all 155,300 app usages, 20,202 (13\%) include at least one text entry event. 

\begin{table}[htbp]
    \small
    \renewcommand{\arraystretch}{1.1}
    \newcolumntype{L}{>{\raggedright\arraybackslash}X}
    \begin{tabularx}{\textwidth}{|X|X|X|X|X|X|}
        \hline
        N & \%Text-Session & Duration-Mean & \#Char/Text-Session & \%TextEntry & \%Structured \\
        \hline
        20, 202 & 13.0\% & 27.9s (SD=80.6s) & 23.6 (SD=51.0) & 5.25\% & 2.50\%\\
        \hline
    \end{tabularx}
    \begin{tablenotes}[normal, flushleft]
    \item[] \textbf{Keys:} 
    \item[] \textbf{\%TextSession:} Percentage of text entering app usages out of all app usages
    \item[] \textbf{DurationMean:} Mean duration for each text entering app usage
    \item[] \textbf{\#Char/TextSession:} Mean number of characters entered for each text entering app usage
    \item[] \textbf{\%TextEntry:} Percentage of \textit{text heavy} app usages (with more than 100 characters entered) in all text entering app usages
    \item[] \textbf{\%Structured:} Percentage of strings containing structured data in all strings entered 
    
    \end{tablenotes}
    \caption{Descriptive Statistics for App Usage}
    \label{tab:descriptive_stats}
\end{table}

In Table~\ref{tab:descriptive_stats}, we show the descriptive statistics for the 20,202 app usages with text-editing events (which we will call text entering app usages). For the analysis, we consider an app usage to be ``text entry heavy'' if more than 100 characters are entered into the app during this app usage. We also parsed the contents of the text entered and consider a piece of text to be ``structured'' if it contains an email address, a URL, a phone number or a US-formatted street address.

One thing we noticed in our dataset is that there are distinct differences in the amounts of text entry among participants. The 5 participants with the smallest number of characters only entered on average 8,731 characters (SD=1519.7) during the study, while the 5 participants at the other end of the spectrum entered an average of 50,253 characters (SD=12789.8). With respect to app usages, the participants also exhibited diverse text entry behaviors: the bottom 5 participants (which are different than for characters entered) on average entered 12 characters (SD=3.5) per text entering app usage, while the top 5 on average entered 51 characters (SD=9.3).  An interesting fact is that the 2 participants with the least characters entered per text entering app usage actually had the largest total number of text entering app usages. This suggests that users exhibit very different text entry patterns, with some entering a large number of shorter strings and some entering fewer, but longer, strings. 

We find no statistically significant difference between the two gender groups of our participants in the average amount of text entered per person (Male 28038.1, Female 28005.5), average number of text entering sessions per person (Male 1280.9, Female 1123.6) or the mean session duration (Male 29.9s, Female 30.9s).

\begin{table}[htbp]
    \small
    \renewcommand{\arraystretch}{1.1}
    \newcolumntype{L}{>{\raggedright\arraybackslash}X}
    \rowcolors{2}{gray!25}{white} 
    \begin{tabularx}{\textwidth}{|l|l|X|}
        \hline
        \rowcolor{gray!50}
        Category & Usage\# & Example Apps \\
        \hline
        Communication-IM & 26,178 & WhatsApp, Facebook Messenger\\
        Tools-Other & 25,173 & Google Search, Clock, Calculator, Dictionary, File Manager\\
        Social & 7,594 & Facebook, Twitter, Pinterest\\
        Communication-Browser & 4,820 & Chrome, Samsung Browser, Dolphin Browser\\
        Communication-Email & 3,633 & Gmail, Inbox by Gmail, Yahoo! Mail\\
        Communication-SMS & 3,446 & Messaging, Verizon Messages\\
        Tools-Dialer & 2,466 & Phone, Dialer\\
        Transportation & 1,302 & Uber, Lyft, RideTrack\\
        News \& Magazine & 996 & BaconReader for Reddit, NYTimes, Flipboard, Feedly\\
        Media \& Video & 963 & Youtube, Gallery, VLC\\
        Music \& Audio & 852 & Spotify, Google Music\\
        Communication-Others & 834 & Contacts, Visual Voice Message, TrueCaller\\
        Travel \& Local-Maps & 744 & Google Maps, Waze, HERE Maps\\
        Productivity-Others & 700 & Calendar, Adobe Reader, IFTTT, Dropbox\\
        Photography & 617 & Camera, Google Photos, Pixlr\\
        Entertainment & 436 & Netflix, 9GAG, IMDb\\
        Productivity-Text Editor & 422 & Google Docs, POLARIS Office, Hamcom Office\\
        Finance & 403 & Square, TurboTax, Venmo\\
        Health \& Fitness & 366 & Fitbit, S Health, Nike+\\
        Lifestyle & 362 & Starbucks, Tinder, Zomato\\
        Productivity-Memo & 324 & Evernote, Keep, Samsung Memo\\
        
        \hline
    \end{tabularx}
    \caption{App Category and Example Apps Sorted by Usage}
    \label{tab:app_category}
\end{table}

\paragraph{RQ1: To which applications are data entered into smartphones?}

To answer RQ1 on understanding where the data are entered into smartphones, we logged the smartphone apps in which each text entry is entered, and group these apps using the categories from the Google Play store. We also further divide the text entry heaviest categories into sub-categories to better characterize the users' text entry behaviors in those apps. In Table~\ref{tab:app_category}, we list the top 20 app categories with the most app usages and example apps within the category. Note that we have excluded system services and launchers from this table. Among all the text entered, 92.6\% of the total 476,322 characters were entered into the top five most text-heavy categories: Communication-IM (64\%), Communication-SMS (17\%), Social (4.6\%), Communication-Browser (4.0\%) and Communication-Email (3.1\%).  All of the above five except for Email are also among the categories with the most number of text entering app usages. Among all the 20,202 text entering app usages, 12,477 (61.8\%) are for IM, 1,588 (7.9\%) are for SMS, 1,185 (5.9\%) are for Tools-Other, 970 (4.8\%) are for Browsers and 822 (4.1\%) are for Social.

\begin{figure}
	\centering
	\includegraphics[width=0.6\columnwidth]{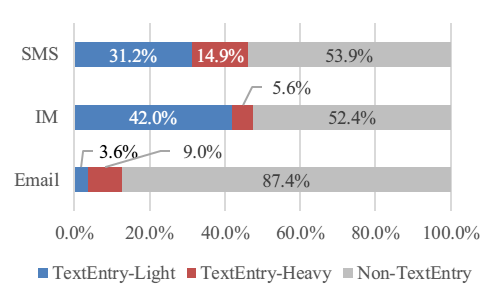}
	\caption{Proportions of Light, Heavy and Non-Text Entry Usage for Three Types of Communication Apps}
	\label{fig:text_entry_propotion}
\end{figure}

An interesting observation is that, within the Communication category, the subcategories exhibit very different profiles for the number of characters entered per text entering usage and the percentage of text entering usages in all usages. As shown in Figure~\ref{fig:text_entry_propotion}, for SMS and IM, 46.1\% and 47.6\% of usages respectively include text entry, but only 12.6\% of email usages involve text entry. Among all the text entering usages, 14.9\% SMS text entering usages are text-entry heavy (having 100+ characters entered), while the same statistic is 9.0\% for email and 5.6\% for IM. 

This result implies that our participants are far less likely to send out or reply to emails on the smartphone compared to SMS and IM messages. This is consistent with what people said in our pre-interviews. When we asked our participants on how they would split tasks between smartphone and other computing devices, all of our participants said that they would not write long emails on their phones unless it was urgent. Instead, they would prefer to wait until they had access to a computer. About half of our participants reported that they seldom write any email on smartphones, mostly because ``\textit{it is hard and frustrating to type long emails on [a] phone.}'' Others said they would reply to emails if it was quick, but would avoid writing formal or important emails because ``\textit{formatting emails on [a] smartphone is hard.}'' We discovered significant differences in the proportions of text heavy usage for SMS apps and IM apps (t=9,48, p\textless.001). As shown in Figure~\ref{fig:text_entry_propotion}, the percentage of text heavy usages is 2.7 times higher in SMS apps than in IM apps, which is interesting as we often consider these two apps to be used in similar scenarios. We asked some of our participants in the exit interview about how they would differentiate the use of SMS and that of IM, and got interesting insights. Three participants mentioned that the availability and easy accessibility of multi-media contents like pictures, video clips, animations (or ``GIFYs''), emojis and audio messages in most IM apps reduce the need for typing long text messages. They said they would avoid typing long messages as much as possible and preferred to send the ``alternative'' form of messages because it was hard to type on phones. However, in situations like when the message recipient does not use IM apps, does not use the same IM app as the user does, or is not a friend on IM with the user, the user would have to use SMS to reach the recipient. This forces the user to type a longer message than they would when using IM apps due to the lack of easily accessible multimedia contents.

Other participants offered different perspectives on the rationale behind the text entry pattern difference between SMS and IM apps. One said that ``\textit{because [WhatsApp] is free, it incurs no extra cost for sending many shorter replying messages like `Yeah', `OK' or `Sure'. But since I'm on a pay-by-use mobile plan, I need to pay extra to send SMS. That's why I send fewer SMS and try to cover more information in one message.}''     

\paragraph{RQ2. What kinds of data are entered into smartphones?}

\begin{figure}
	\centering
	\includegraphics[width=0.6\columnwidth]{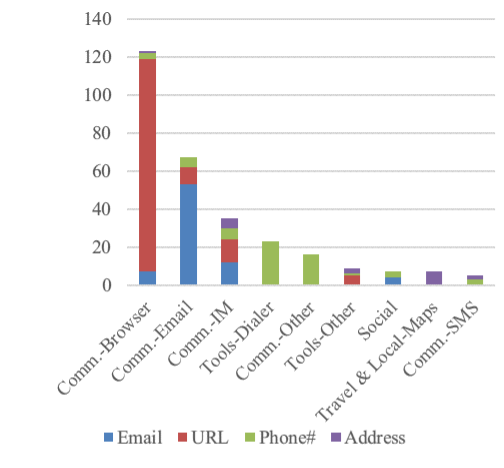}
	\caption{Distribution of Structured Data Entry Usage across App Categories (“Comm” is the Communications category)}
	\label{fig:text_entry_distribution}
\end{figure}

To investigate RQ2, we parsed and analyzed all 12,504 strings entered into the phones by our participants during the course of the study. We looked for structured data, including URLs, email addresses, phone numbers and street addresses. We were able to find 312 instances of strings with structured data, among which 72.8\% include URLs, 27.6\% include email addresses, 19.6\% include phone numbers and 7.7\% include street addresses. Note that the sum of counts for different types of structured data can exceed the total number of strings with structured data, as a string can contain more than one type of structured data, but we only count once if same type of structured data shows up more than once in a single string. We show the distribution of structured data entry usage among application categories in Figure~\ref{fig:text_entry_distribution} for categories with at least 5 strings with structured data entered. 

This result confirms the hypothesis that the type of data entered is associated with the nature of the app~\cite{chang_improving_2013}. Different types of structured text go to where you might think they should go---the participants entered URLs into browsers, email addresses into email clients, phone numbers into contacts and dialer apps, and street addresses into maps. 

However, if we compare the amount of structured data entered with the number of total text entering usages for each category of app, the portion of text entering usages with structured data entered is surprisingly small. As an example, for the browsers, only 12.2\% of the text entering sessions had a URL entered. Similarly, only 11.6\% text entering sessions for email apps had an email address entered, 15.6\% text entering sessions for dialer had a phone number entered and 5\% text entering sessions for maps had a street address entered.

Talking with the participants in the exit interview, most of the participants reported that they seldom actually type structured data into any apps. The text fields in which one might expect structured data to be typed the most---the ``To'' fields in email clients, number fields in dialers, address bars in browsers or the address fields in map apps---are actually used as the de facto search bars in many modern app design, where the user can just type in the name (and in most cases, the partial name) of the contacts, web-sites or points of interest instead of the corresponding numbers, URLs or addresses. This technique reduced the user's typing workload and eliminated many needs for typing structured data. 

Based on the results shown in Figure 2, for example, it is still safe to assume that users are more likely to type email addresses into email clients than into any other kinds of app, and to assume that users in email clients are more likely to type an email address than other types of structured data. But it would be wrong to assume that the user will type an email address most of the time when they send emails. The same goes for other combinations of structured data types and apps.

\paragraph{RQ3. Through what means are data entered?}

We then looked at RQ3 on how the data are entered into the phones. Three major means of data entry were considered: typing, copy-and-paste and auto-completion. A prior study found that not presenting the auto-completion suggestions enabled the users to type faster, but the suggestions were utilized to save taps and were subjectively preferred~\cite{quinn_cost-benefit_2016}. Copy-and-paste was also perceived to be difficult by users in a prior study~\cite{chen_bezelcopy:_2014}. 

In our 401,875 text-editing events collected, 51,534  (12.8\%) were deletions, 311,369 (77.5\%) were additions, in which 285,772 events were typing, 158 events were copy-and-pastes and 25,439 events were auto-completion. (For typing, each keystroke is a text-editing event. Each instance of paste or auto-completing is also an event.) The 12.8\% of text entry events being deletions seems higher than other studies (e.g., 9.7\% in~\cite{wobbrock_analyzing_2006} and 9.5\% in~\cite{mackenzie_text_2002}). Lengthwise, the 285,772 characters typed constituted 60.0\% of the total 476,322 characters entered. 12,730 (2.7\%) characters were pasted in, and the rest, 177,870 (37.3\%), were entered using auto-completion or using a suggestion. 

\begin{figure}
	\centering
	\includegraphics[width=0.6\columnwidth]{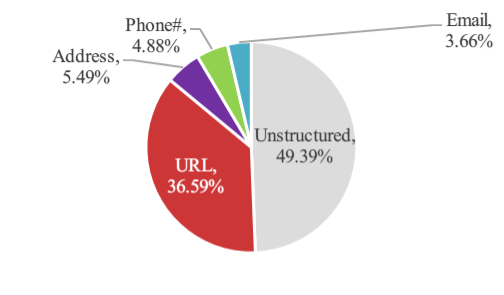}
	\caption{Clipboard Contents by Structure Type}
	\label{fig:text_clipboard}
\end{figure}

We investigated the characteristics of the content being copied into the clipboard. During the course of the study, 164 unique items were copied into the clipboard across all participants. That is only an average of 0.69 items per participant per day.   

Among the texts that were copied, 50.6\% contained structured data, among which the majority were URL links (72.3\%). The distribution of content type for clipboard content is shown in Figure 3.

The low amount of clipboard usage confirms what our participants said in the pre-survey: that they do not use copy-and-paste often, as they find it frustrating. The participants also expressed the same frustration in the exit interview. When asked about the alternative means for transferring data between fields other than using copy-and-paste, we heard three major methods mentioned. For text that is short and easy to remember, many would simply retype. For sharing longer quotes, tweets or social network statuses with friends, a few of our participants mentioned that they often just made a screenshot and sent the picture to their friends via IM apps. They considered using this technique to be quicker and easier, as taking a screenshot only requires pressing the power button and the volume-down button simultaneously---which is much easier to do than selecting the right block of text with one's fingers on the small screen. Text sent through screenshot can also be easier for the recipient to read as it preserves the format of the content.

In the exit interview, most users identified URLs as their most commonly copied structured data, which is consistent with our observations in the data. A participant reflected ``\textit{if I am to share a web content with my friends, I would normally just send a screenshot. But sometimes I'll copy-and-paste the link to the chat instead if the page is too long. Sometimes I'll also have to copy a link to my browser if it is not clickable.}''

\begin{table}[htbp]
    \small
    \renewcommand{\arraystretch}{1.1}
    \newcolumntype{L}{>{\raggedright\arraybackslash}X}
    \begin{tabularx}{\textwidth}{|l|X|X|X|X|X|}
        \hline
        N & \#AppUsage-Mean & Duration-Mean & \#UniqueApp-Mean & \%Text-Entering & \%MultiApp \\
        \hline
        17,684 & 4.76 (SD=8.64) & 181.7s (SD=404.1s) & 1.99 (SD=1.23) & 39.9\% & 55.6\%\\
        \hline
    \end{tabularx}
    \begin{tablenotes}[normal, flushleft]
    \item[] \textbf{Keys:} 
    \item[] \textbf{\#AppUsageMean:} Average number of non-system app usages in sessions
    \item[] \textbf{DurationMean:} Mean duration for each session
    \item[] \textbf{\#UniqueAppMean:} Average number of unique non-system apps used in sessions 
    \item[] \textbf{\%TextEntering:} Percentage of sessions with text entry
    \item[] \textbf{\%MultiApp:} Percentage of sessions containing more than one app usages 
    
    \end{tablenotes}
    \caption{Descriptive Statistics for Sessions}
    \label{tab:descriptive_stats_sessions}
\end{table}

\subsection{Text Entry in Cross-application Interactions}

In this section, instead of looking at the usage of individual apps, we focus on understanding text entry in the context of multi-app sessions. To characterize the user's text entry behavior in multi-app sessions, we investigated the following two research questions:

\begin{itemize}

\item[\textbf{RQ4:}] How are text-entering sessions different from the non-text entering sessions?

\item[\textbf{RQ5:}] How is text transferred among apps within sessions?

\end{itemize}

As mentioned above, we define a \textit{session} as containing all app usages from when the device is powered on or awakened until when the device is powered off or turned to standby mode. Based on this, we group 155,300 app usages in our dataset into 37,496 sessions.  This definition of sessions is reasonable based on previous findings~\cite{van_berkel_systematic_2016} that for the majority of instances where users return to their smartphone i.e., unlock their device, they in fact begin a new task as opposed to continuing a previous one. Among our 37,496 sessions, 52.8\% are turning on the phone display without launching any non-system apps, which are mostly to check the time or to see the notifications on the lock screen. We exclude those sessions in the following analyses. In Table~\ref{tab:descriptive_stats_sessions}, we show the descriptive statistics for the 17,684 sessions with at least one app usage.

The statistics confirm phenomena reported by prior work. In Jesdabodi and Maalej's study~\cite{jesdabodi_understanding_2015}, which used the LiveLab dataset~\cite{shepard_livelab:_2011}, they reported a mean of 2.12 unique apps per session and overall mean duration of usage sessions of 172.8s, which are similar to our stats. We also find 58.9\% of our sessions last less than 30s and 87.0\% last less than 4 minutes, which supports Yan et al.'s finding~\cite{yan_fast_2012} that 50\% of phone engagements last less than 30 seconds, and 90\% last less than 4 minutes. Our descriptive stats of sessions are also in line with Carrascal and Church's prior study~\cite{carrascal_-situ_2015} and confirms Ferreira et al.'s finding about mobile application micro usage~\cite{ferreira_contextual_2014}.  

\begin{table}[htbp]
    \small
    \renewcommand{\arraystretch}{1.1}
    \newcolumntype{L}{>{\raggedright\arraybackslash}X}
    \begin{tabularx}{\textwidth}{|l|l|X|X|X|X|}
        \hline
         & N & \#AppUsage-Mean & Duration-Mean & \#UniqueApp-Mean & \%MultiApp \\
        \hline
        Non-Text & 10,630 & 2.39 (SD=2.45) & 129.7s (SD=339.9s) & 1.66 (SD=0.93) & 43.2\%\\
        Text & 7,054 & 8.34 (SD=12.5) & 260.0s (SD=474.4s) & 2.49 (SD=1.44) & 74.4\%\\

        \hline
    \end{tabularx}

    \caption{Comparing Descriptive Statistics for Non-Text Entering Sessions and Text-Entering Sessions}
    \label{tab:compare_nontext_test}
\end{table}

\paragraph{RQ4: How are text-entering sessions different from the non-text entering sessions?}

Table~\ref{tab:compare_nontext_test} shows a comparison of the descriptive statistics between the text entry sessions and the non-text entry sessions. We see that text entry sessions have higher average number of non-system app usages per session (t=39.5, p\textless0.001), longer mean duration (t=19.9, p\textless0.001), higher average number of unique apps in the sessions (t=42.8, p\textless0.001) and higher percentage of multi-app sessions (t=44.1, p\textless0.001). 

\paragraph{RQ5: How is text transferred among apps within sessions?}

To answer RQ5, we consider two methods of data transfer between apps: copy-and-paste and the Android built-in ``Share to'' API. 

One of the most obvious ways to transfer data between apps is through copy-and-paste. A prior study suggested that at least half of copy-paste operations were cross-app~\cite{chen_bezelcopy:_2014}. We discussed above the content of the clipboard and the low usage of copy-and-paste in RQ3 of the previous section. Here we focus on where are data copied from and to.

In Figure 4, we show the distribution of 164 copy events (including cuts) and 158 paste events across the app categories. 41\% of the paste events are cross app, which is defined as that the source of the clipboard content is different from the destination for pasting.  The vast majority of copy-and-paste events (84.6\%) are in the communication category, which is not surprising giving the dominance of communication app in the number of text entry sessions. The structure type of the pasted content is also associated with the app category. For instance, 67\% of strings pasted into browsers include URL links, and 78\% of strings with URLs in the clipboard are pasted into a browser.

Another method of data sharing across apps is through the Android ``Share to'' mechanism, which allows the user to share the current content to a set of apps that can ``consume'' this type of content. The developers for both the source app and the destination app have to explicitly specify the content of the data to share, the type of the data to share and the type of the data to accept. This mechanism is popular among our participants and covers many of the most frequent use cases like sharing a web page to WhatsApp chat, a piece of text to Gmail, etc.

\begin{figure}
	\centering
	\includegraphics[width=0.6\columnwidth]{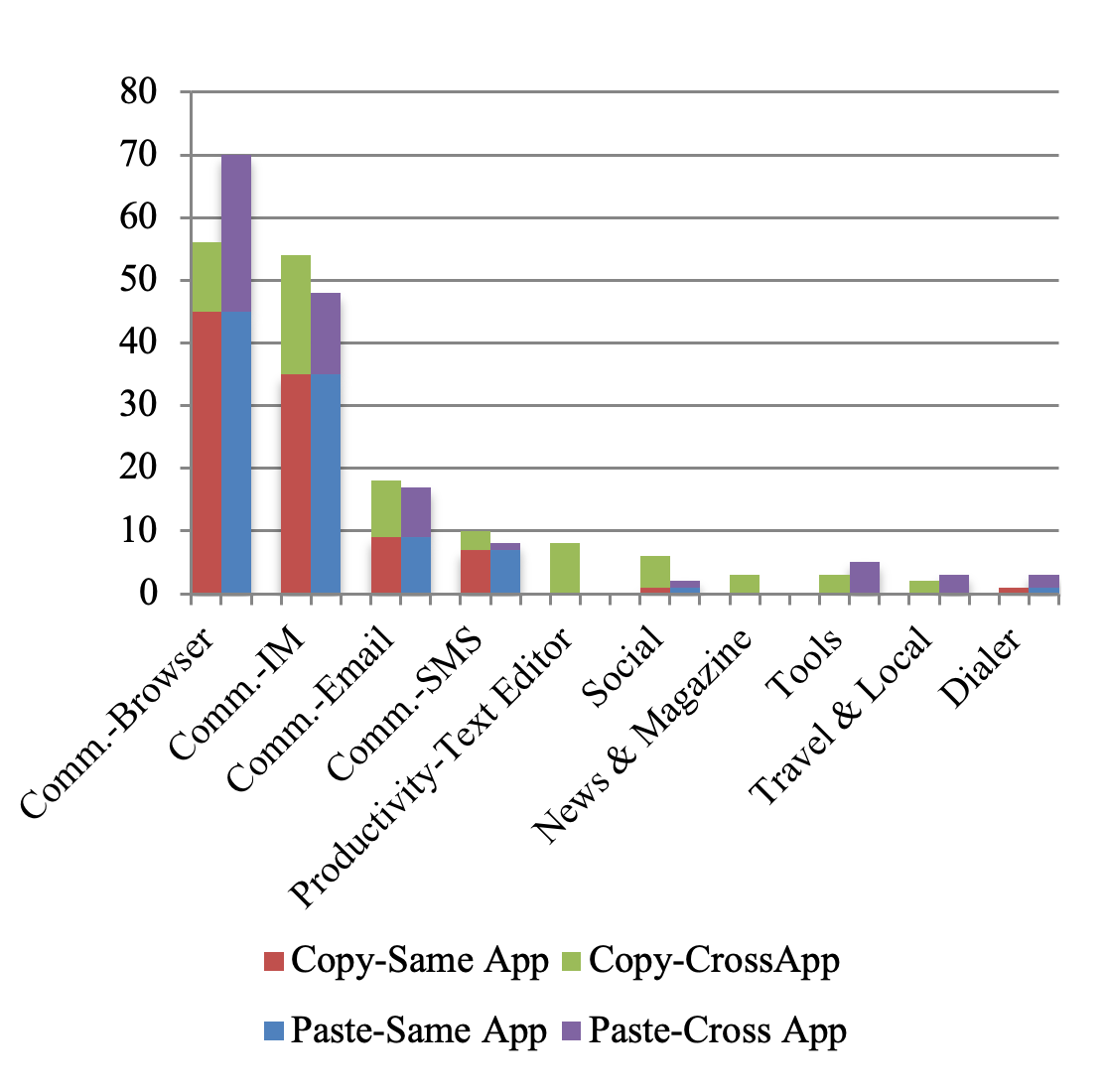}
	\caption{Distribution of Copy-and-Paste Counts across App Categories. “Copy-Same” means the copy or cut operation was in an app, and the corresponding paste was also in the same app. “Copy-CrossApp” means the copy/cut was in an app and the paste was in a different app. Similarly for “Paste-xx” where the paste was in this category and the copy/cut was either here or in a different app.}
	\label{fig:text_entry_copy_pate_distribution}
\end{figure}

We were not able to record the number of uses of the ``share to'' mechanism, since many mobile apps have overridden the default ``share to'' API provided by the Android operating system. However, in the interviews, our participants mentioned they often use this mechanism since ``\textit{it better fits my way of thinking about the task. It's also much easier to use than copy-and-paste}''. 

\section{Discussion}
In the results, we see fewer instances of structured data entered than one might expect. From the interviews, we learn that this is mainly because in current app design, data entry fields can also be used as de facto search bars where the user can simply enter a partial entity name (e.g., the website name or recipient name) and the structured data (e.g., the URL or email address) will be retrieved. Such design is well used and liked by our participants.

Similarly, there are only a small number of copy-and-paste instances in our dataset. In interviews, the participants expressed their frustration about the current copy-and-paste mechanism, and talked about how they adopt alternative methods to transfer data between apps through the ``screen-shot-and-send-picture'' method, the Android ``share to'' mechanism, or app-specific ways of sharing data (e.g., through cloud service). This suggests that higher level support is needed to facilitate the data transfer between apps in cross-app tasks. For sharing web content with friends, the design for better supporting ``sharing beyond the blue link'', as framed in~\cite{bentley_i_2016}, is also worth exploring.  

Our results support the need for better integration and more streamlined data flow among apps. Users prefer to not type data into fields or copy-and-paste text between fields. Thus, they quickly adopt interactions techniques like integrated search and intent-based cross-app data sharing mechanisms. So, in the design of mobile interfaces, the designers should minimize the unnecessary or repetitive data entry, and support automatically sharing data from one app to another. 

The results also show a correlation between the category of app in use and the type of structured data entered. This implies that predictive text entry methods should not only consider the language model and the semantic context, but also the context for the app usage. 

\section{Conclusion and Future Work}
We conducted a mixed-method study on understanding user's text entry on smartphones in the context of app usage. Our results fill in gaps in prior work, identify characteristics of users' current text entry usage, and shed light on requirements for new systems. We characterize mobile text entry as to what, how, and to which apps the contents are entered, based on actual usage data collected from our participants. We also consider text entry in the context of cross-app sessions and compare text-entering and non-text-entering sessions.

In this work, we studied the usage by a sub-population of smartphone users and provide insights on designing novel interactions to better support their usage. Future research might investigate text entry behaviors for populations with broader demographics, especially those from diverse socioeconomic backgrounds, people in developing countries~\cite{pew_research_center_communications_nodate}, or field workers who have limited access to computers and use smartphone as their primary computing device~\cite{karlson_working_2009}. Per our interviews, our participants seldom use voice input, but a future study should collect voice input data and compare how the usage, scenario and context of voice input differ from those of other text entry methods. It would also be interesting to investigate a broader range of text entry, like text entry across multiple devices. 

Another needed study would be about the relationships between data entry and the user's task flow, and investigate what we can infer about a user's other activities based on their text entry behaviors, and vice versa.  Finally, the resulting knowledge could guide the design of novel systems and interaction techniques to better support users' tasks on smartphones.

\bibliographystyle{SIGCHI-Reference-Format}
\bibliography{references}

\end{document}